\documentclass[twocolumn, times]{aastex63}
\graphicspath{{./}{figures/}}
\usepackage{amsmath}

\makeatletter
\def\coldhead{coldhead}
\makeatother

\newcommand{\ms}[1]{$#1~{\rm m\,s^{-1}}$}
\newcommand{\todo}[1]{\textcolor{blue}{[#1]}}
\newcommand{\citeme}[1]{\todo{citations}}

\received{July 10, 2019}
\revised{September 16, 2019}
\accepted{October 2, 2019}
\submitjournal{ApJL}

\begin{document}

\title{Spiral Structure in the Gas Disk of TW Hya}

\correspondingauthor{Richard Teague}
\email{rteague@umich.edu}

\author[0000-0002-0786-7307]{Richard Teague}
\affil{Department of Astronomy, University of Michigan, 311 West Hall, 1085 S. University Ave, Ann Arbor, MI 48109, USA}

\author[0000-0001-7258-770X]{Jaehan Bae}
\affil{Department of Terrestrial Magnetism, Carnegie Institution for Science, 5241 Broad Branch Road NW, Washington, DC 20015, USA}

\author[0000-0001-6947-6072]{Jane Huang}
\affil{Harvard-Smithsonian Center for Astrophysics, 60 Garden Street, Cambridge, MA 02138, USA}

\author[0000-0003-4179-6394]{Edwin A. Bergin}
\affil{Department of Astronomy, University of Michigan, 311 West Hall, 1085 S. University Ave, Ann Arbor, MI 48109, USA}

\begin{abstract}
We report the detection of spiral substructure in both the gas velocity and temperature structure of the disk around TW~Hya, suggestive of planet-disk interactions with an unseen planet. Perturbations from Keplerian rotation tracing out a spiral pattern are observed in the SE of the disk, while significant azimuthal perturbations in the gas temperature are seen in the outer disk, outside 90~au, extending the full azimuth of the disk. The deviation in velocity is either $\Delta v_{\phi} \, / \, v_{\rm kep} \sim 0.1$ or $\Delta v_{z} \, / \, v_{\rm kep} \sim 0.01$ depending on whether the perturbation is in the rotational or vertical direction, while radial perturbations can be ruled out. Deviations in the gas temperature are $\pm 4$~K about the azimuthally averaged profile, equivalent to deviations of $\Delta T_{\rm gas} \, / \, T_{\rm gas} \sim 0.05$. Assuming all three structures can be described by an Archimedean spiral, measurements of the pitch angles of both velocity and temperature spirals show a radially decreasing trend for all three, ranging from 9\degr{} at 70~au, dropping to 3\degr{} at 200~au. Such low pitch-angled spirals are not readily explained through the wake of an embedded planet in the location of previously reported at 94~au, but rather require a launching mechanism which results in much more tightly wound spirals. Molecular emission tracing distinct heights in the disk is required to accurately distinguish between spiral launching mechanisms.
\end{abstract}

\keywords{Circumstellar disks -- Planet formation -- Protoplanetary disks -- Interferometry}

\section{Introduction}
\label{sec:intro}

It appears that substructure is ubiquitous in protoplanetary disks, particularly in the dust distributions \citep[e.g.][]{Andrews_ea_2018, Avenhaus_ea_2018}. Various mechanisms have been shown to reproduce this structure, including the presence of unseen protoplanets \citep{Zhang_2018}, and various (magneto-)hydrodynamical instabilities \citep{Flock_ea_2015}. Small grains, however, are a relatively passive component of the disk and are readily shepherded by the gas with their dynamics dictated by the gas pressure gradients \citep{Whipple_1972}. Therefore, to understand and distinguish between the dynamical processes shaping the dust structure we must look to substructures in the gas.

In particular, gas dynamics have been shown to reveal a large variety of substructure previously undetected. For example, \citet{Teague_ea_2018a, Teague_ea_2018c} demonstrated that slight deviations in the rotation velocity, $v_{\phi}$, were detectable, indicative of local changes to the gas pressure gradient, and thus allowing for the inference of the true underlying gas density profile. Local deviations in the velocity structure have also been detected. \citet{Pinte_ea_2018b} reported the detection of a `kink' in an iso-velocity contour in HD~163296, arguing that an embedded planet of $\sim 2~M_{\rm Jup}$ was required to distort the gas rotation to such an extent. More recently, \citet{Pinte_ea_2019} reported a similar feature in the disk of HD~97048. Similarly, \citet{Casassus_Perez_2019} showed evidence of an embedded planet in HD~100546 through a `Doppler-flip` in the residuals between the rotation map and a best-fit azimuthally averaged velocity structure, confirming the predictions by \citet{Perez_ea_2018}.

Here we focus on the disk of TW~Hya, the closest protoplanetary disk at $60.1~\pm0.1$~pc \citep{Bailer-Jones_ea_2018}. Significant sub-structure is observed in both the mm-sized grains \citep{Andrews_ea_2016, Nomura_ea_2016, Huang_ea_2018} and sub-\micron{} grains \citep{Debes_ea_2013, Debes_ea_2016, Rapson_ea_2015, vanBoekel_ea_2017}. Both $^{12}$CO and CS have been observed to have similar dip features at these locations in their emission profiles \citep{Teague_ea_2017, Huang_ea_2018}, where later \citet{Teague_ea_2018b} demonstrated through a non-LTE excitation analysis that there was a significant drop in the column density of CS at 90~au, suggestive of a significant depletion in the total gas surface density.

In this Letter we search for comparable azimuthal structure in the gas physical structure, both in terms of gas velocity and temperature. In \S\ref{sec:observations} we describe the observations and the methods use to extract this structure, characterising the spirals and discussing potential launching mechanisms in \S\ref{sec:discussion}. We conclude with a summary in \S\ref{sec:summary}.

\section{Observations}
\label{sec:observations}

We use the $^{12}$CO $J = 3-2$ observations presented in \citet{Huang_ea_2018} which make use of the ALMA projects 2015.1.00686.S (PI: Andrews) and 2016.1.00629.S (PI: Cleeves). The final image has a beam size of $139~{\rm mas} \times 131~{\rm mas}$ with a position angle of $105\degr$ and a channel spacing of \ms{250}. We refer the reader to the aforementioned publication for full details of the data reduction and subsequent imaging.

$^{12}$CO emission offers an excellent probe of the disk physical structure. As it is optically thick, the line intensity provides an accurate measure of the local gas temperature at the $\tau \approx 1$ emission surface. In addition, low-$J$ transitions are exceptionally bright, allowing for an accurate determination of the line center, and thus velocity structure, across the entire 200~au radius of the disk.

\subsection{Velocity Structure}
\label{sec:v0structure}

\begin{figure*}
    \centering
    \includegraphics[width=\textwidth]{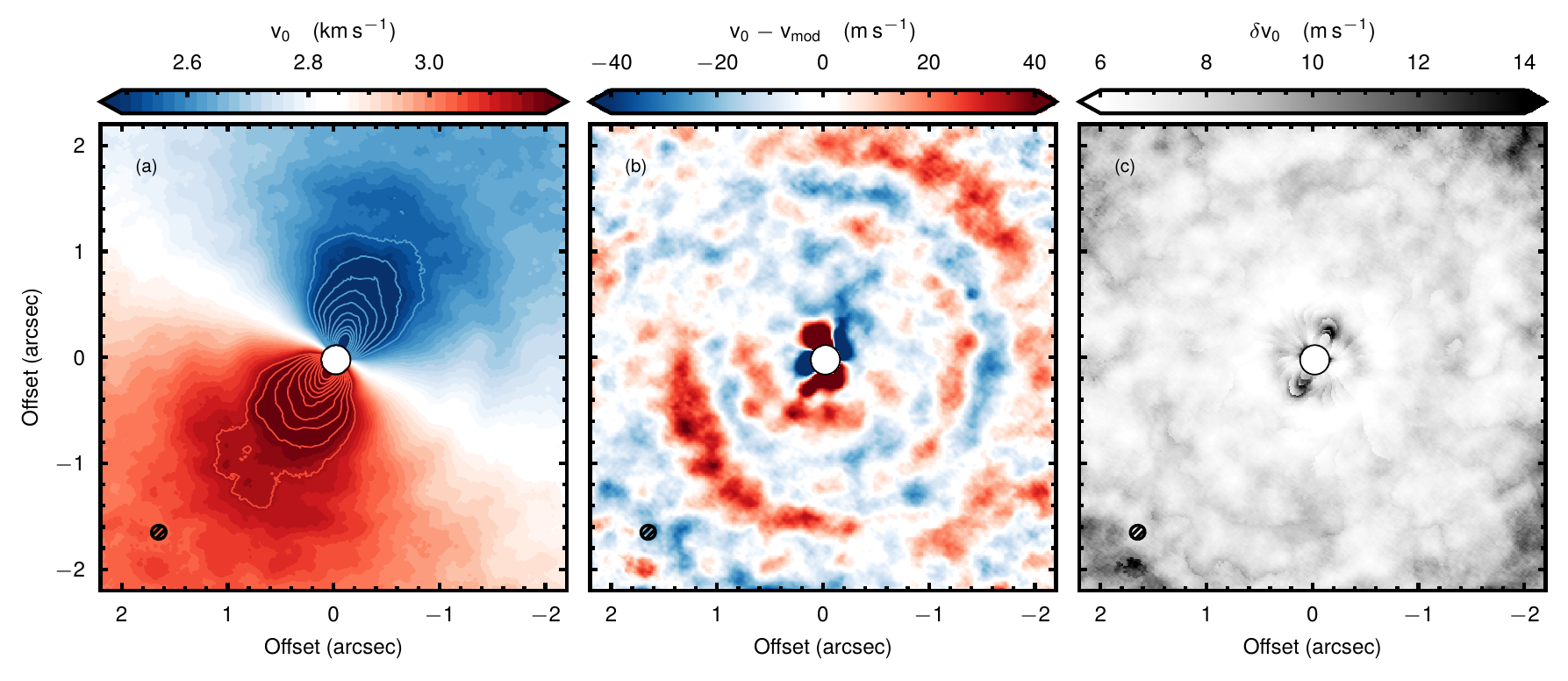}
    \caption{\textbf{(a)} Rotation map of the $J = 3-2$ emission. High velocity offsets are clipped in the colour scaling to emphasise the structure when $v_0 \sim v_{\rm LSR}$. The line contours show steps of \ms{50}. The synthesised beam is shown in the bottom left of each panel. \textbf{(b)} Residuals from the rotation map with the best-fit Keplerian rotation model. \textbf{(c)} Uncertainties used for the fitting of the rotation map, derived using \texttt{bettermoments}.}
    \label{fig:v0_structure}
\end{figure*}

To make the rotation map we, use the \texttt{bettermoments}\footnote{\url{https://github.com/richteague/bettermoments}} package using the quadratic method described in \citet{Teague_Foreman-Mackey_2018} to measure the line centers, $v_0$. This approach fits a quadratic curve to the pixel of peak intensity and the two neighbouring pixels. This provides an excellent measure of the line centroid when the emission is dominated by a single component and allows for a level of precision much greater than the velocity resolution of the data.

As the channel spacing of the data is comparable to the line width of the $^{12}$CO emission, the systematic effect of the spectral response function, namely a broadening of the line profile, will be significant \citep[see, for example,][]{Koch_ea_2018}. To mitigate these effects, we follow the approach described in \citet{Christiaens_ea_2014} and image the data at two additional velocity offsets, $\pm$\ms{100}. The line center was calculated for each of these images, then averaged and the uncertainties combined. A similar approach was used to make the $T_{\rm B}$ map in \citet{Huang_ea_2018}.

The resulting map is presented in the left panel of Fig.~\ref{fig:v0_structure}. Already there are clear deviations noticeable along the minor axis of the disk with several `finger' like structures. The uncertainties are plotted in the right-most panel of the same figure, showing that we achieve a precision of less than 10\% of the channel width (\ms{250}).

To highlight the non-Keplerian structure in the $v_0$ map, we use the Python package \texttt{eddy}\footnote{\url{https://github.com/richteague/eddy}} \citep{eddy} to fit a Keplerian rotation profile to the $v_0$ map where the line center at a given pixel is given by,

\begin{equation}
    v_0(r, \, \phi) = \sqrt{\frac{G M_{\rm star}}{r}} \cdot \cos \phi \cdot \sin i + v_{\rm LSR}
    \label{eq:vphi}
\end{equation}

\noindent where $i$ is the inclination of the disk, $v_{\rm LSR}$ is the systemic velocity and $(r,\, \phi)$ are the deprojected cylindrical coordinates. As TW~Hya is so close to face on, it is hard to constrain the inclination and thus we fix it at $5.8\degr$ based on trial runs. A good model of $v_0$, was found with posterior distributions spanning $M_{\rm star} = 0.81_{-0.17}^{+0.16}~M_{\sun}$, ${\rm PA} = 151.6 \pm 5.8~\degr$ and $v_{\rm LSR} = 2839 \pm 18~{\rm m\,s^{-1}}$, consistent with previous constraints from the continuum \citep{Andrews_ea_2018, Huang_ea_2018}. A thorough description of the fitting procedure can be found in Appendix~\ref{sec:app:rotationmapfitting}.

We take the average of 250 random samples of the posterior distributions to create a model rotation map, $v_{\rm mod}$. The residuals when subtracting this model from the observations are shown in the center panel of Fig.~\ref{fig:v0_structure}. Significant residuals are seen which span large regions of the disk, in particular a spiral like feature in the south-east, and an arc structure coincident with the `fingers' in the $v_0$ map. These features will be discussed more in \S\,\ref{sec:discussion}.

\begin{figure*}
    \centering
    \includegraphics[width=\textwidth]{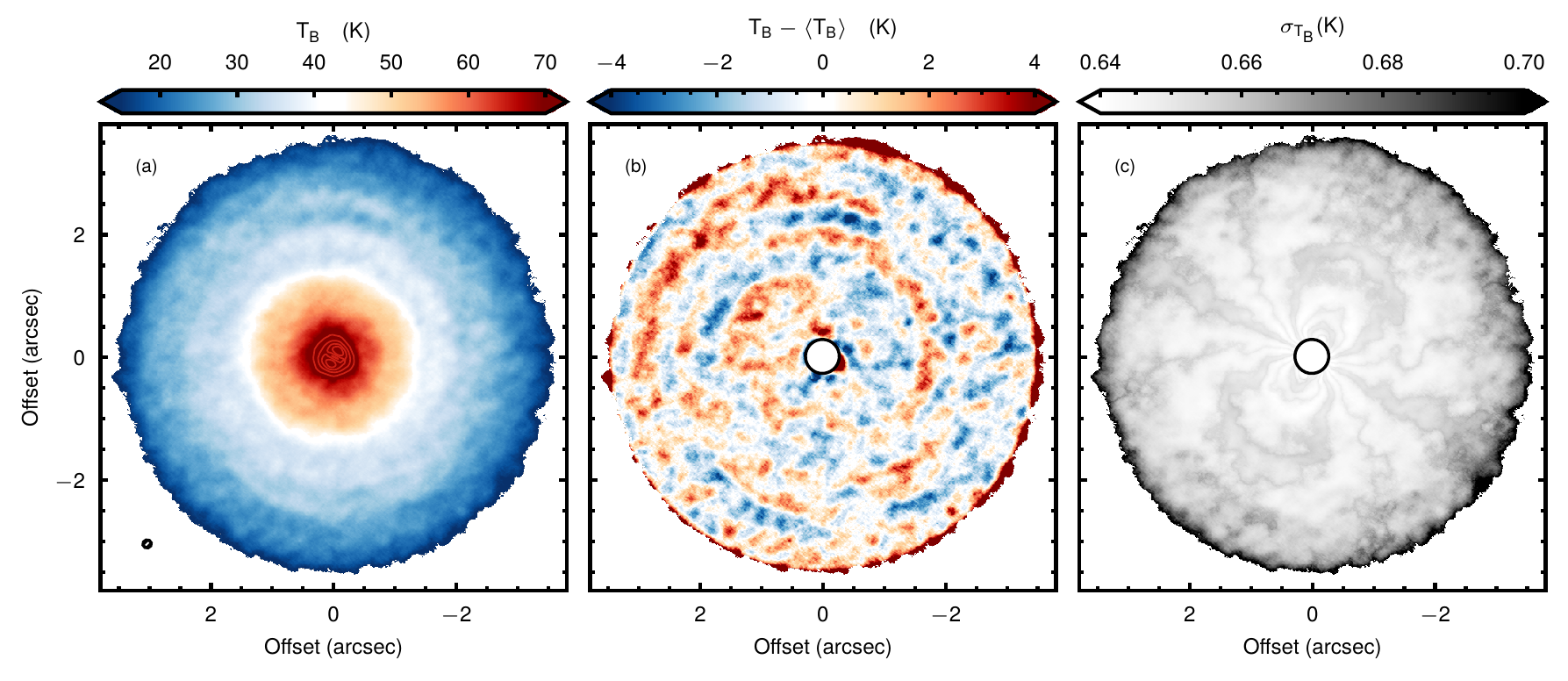}
    \caption{\textbf{(a)} Map of the $^{12}$CO brightness temperature, a proxy of the local gas temperature. The filled contours are clipped at 70~K to highlight structure in the outer disk. The contours start at 80~K and increase in steps of 20~K with a peak $T_{\rm B}$ of 160.7~K. \textbf{(b)} Residuals after subtracting an azimuthally averaged radial profile. The inner 0.4\arcsec{} is masked to hide significant residuals due to beam smearing. \textbf{(c)} Uncertainties in the line peak showing heightened values in regions where the line center is close to the channel edges.}
    \label{fig:Tb_structure}
\end{figure*}

We have additionally attempted the method proposed in \citet{Huang_ea_2018_hydro} to explore the residuals by subtracting one side of the rotation map with the other, however as the structure we observed extends over substantial regions of this the disk, we do not find any improvement over the subtraction of a full 2D model.

\subsection{Temperature Structure}
\label{sec:Tbstructure}

As the $^{12}$CO emission is optically thick, $\tau \gg 1$, we can use the brightness temperature as a probe of the gas temperature. Using the peak values from \texttt{bettermoments}, including the averaging with the spectrally offset images, we convert from flux density units to units of Kelvin assuming the Planck law. The resulting map is shown in the left panel of Fig.~\ref{fig:Tb_structure}, consistent with Fig.~4 in \citet{Huang_ea_2018}, with the associated uncertainties in the right-most panel. More obviously than with the line center maps, the uncertainty varies across the face of the disk due to the spatial dependence on the impact of the spectral response function. As demonstrated in \citet{Koch_ea_2018}, locations where the line center falls at the channel edge will be more strongly affected, resulting in broader line profiles and slightly increased uncertainties. This is why the typical `butterfly' pattern is clearly visible in Figure~\ref{fig:Tb_structure}c.

In a similar approach to the velocity, we subtract an azimuthally averaged radial profile to search for structure in the residuals \citep[see also][]{Cleeves_ea_2015}, which is shown in the middle panel of Fig.~\ref{fig:Tb_structure}. To calculate the radial profile, we deproject the $T_{\rm B}$ map using the inclination and position angle used in the previous section. The recovered radial profile is again consistent with that presented in \citet{Huang_ea_2018} featuring a break in emission at 0.4\arcsec{}, a `shoulder' at 1.2\arcsec{} and a second break at 1.5\arcsec{}.

Subtracting this azimuthally symmetric model leaves significant structure in the residuals, as shown in the central panel of Fig.~\ref{fig:Tb_structure}. Clear spiral structure are observed across the bulk of the disk. As with the residuals in the $v_0$ map, although the deviations are only at a $\sim 5\sigma$ significance in a single pixel, their coherence over large spatial extents points towards real structures.

\subsection{Spiral Structure}
\label{sec:observations:spirals}

There are three clear spirals observed in the residuals: one spanning between a PA of 90\degr{} and 180\degr{} at a radius of 1.3\arcsec{} (78~au) and 1.7\arcsec{} (102~au) in the velocity residuals, and two in the gas temperature, starting at 1.5\arcsec{} (90~au), extending out to 3\arcsec{} (180~au) and covering the full azimuth of the disk. All three spirals have the same orientation, suggesting a clockwise rotation of the disk if the spirals are believed to be trailing, opposite to the counter-clockwise direction inferred by \citet{Debes_ea_2017} using the motion of shadows in the outer disk.

\begin{figure*}
    \centering
    \includegraphics[width=\textwidth]{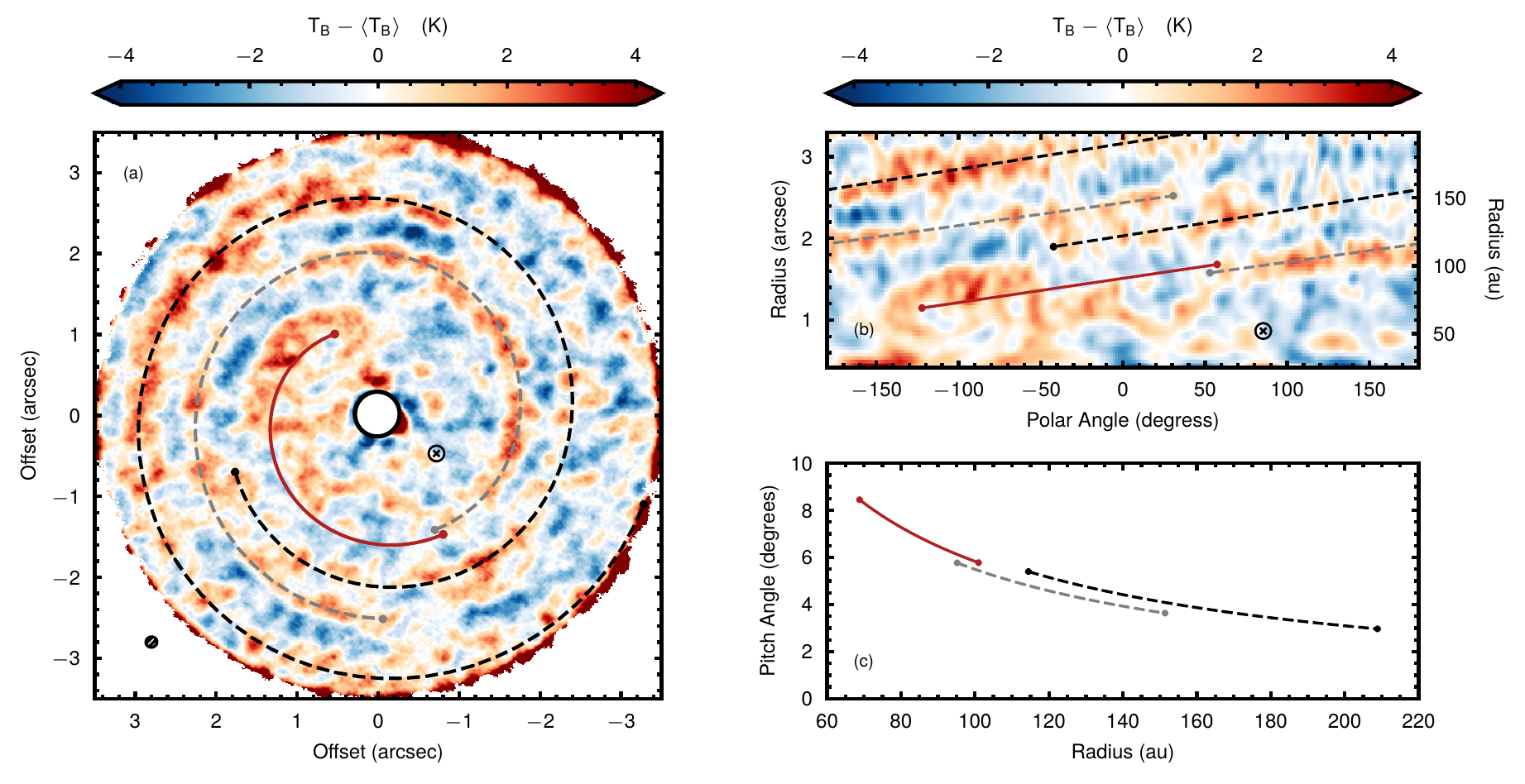}
    \caption{\textbf{(a)} Residuals between the measured $T_B$ and the azimuthally averaged profile, $\langle T_B \rangle$. Linear spirals are overlaid in the dashed lines. The linear spiral in the $v_0$ residuals, as shwon in Fig.~\ref{fig:v0_residuals}, is also overlaid in the solid line. The beamsize is shown in the bottom left of the panel. \textbf{(b)} Polar deprojection of the left panel, clearly demonstrating the linear nature of the spirals. \textbf{(c)} Derived pitch angles for the three spirals. In panels \emph{(a)} and \emph{(b)} the crossed region shows the location of the continuum excess reported by \citet{Tsukagoshi_ea_2019}.}
    \label{fig:Tb_residuals}
\end{figure*}

A spiral shadow has been detected in the scattered NIR which extends between 170\degr{} and 300\degr{} at a radius of $\sim 2\arcsec$ \citep[120~au;][]{vanBoekel_ea_2017}. The shadowed nature suggest either a local decrease in scale height, and thus less incident stellar photons, or a drop in the scattering efficiency. This feature lies between the two warmer spirals, suggesting a region with cooler gas and thus that this is likely either due to a shadow cast from the inner disk or a local scale height depression. A drop in the scattering efficiency would result in more efficient absorption of stellar photons and thus a warmer atmosphere, inconsistent with the raised temperature traced by the $^{12}$CO.

To each of the three spirals we fit an Archimedean, or linear, spiral, $r_{\rm spiral} = a + b \, \phi_{\rm spiral}$. Due to the relatively low significance of the spirals in any given pixel, the fits were performed by eye. The velocity spiral was well fit with $\{a,\,b\} = \{1.51,\, 0.17\}$, while the inner and outer spirals in temperature were characterised by $\{1.43,\, 0.16\}$ ad $\{2.03,\, 0.18\}$, respectively. The spiral fits are shown in Fig.~\ref{fig:Tb_residuals} with the dashed lines showing the spirals in $T_B$, while the solid line shows the spiral in $v_0$, also shown in Fig.~\ref{fig:v0_residuals}. The polar deprojection in Fig.~\ref{fig:Tb_residuals}b highlights the linear nature of the spirals which extend over the bulk of the disk.

Figure~\ref{fig:Tb_residuals}c shows the pitch angles for the three linear spirals given by $\tan \beta = -({\rm d}r / {\rm d}\phi) \, / \, r$. All three show relatively small pitch angles, ranging between 9\degr{} and 3\degr{}. Comparison with the models in \citet{Bae_Zhu_2018b} suggest that for these pitch angles to be driven by a Lindblad resonance requires both a very low scale height, $(h / r)_p \lesssim 0.05$, and perturber much closer in than the $\approx 90$~au gap. This scenario is explored in more detail in the following section. It is also plausible that the two spirals we have identified are in fact a single, broader structure. If this were the case, it would not significantly change the conclusion that these azimuthal structures require \emph{small} pitch angles.

In addition, \citet{Juhasz_Rosotti_2018} showed that in a disk with a positive vertical temperature gradient, the pitch angles will increase with height in the disk owing to the increasing temperature and thus sound speed. Simulations suggested that spirals traced at a $z / r \approx 0.3$, where we expect the $^{12}$CO $\tau \approx 1$ surface to lie, would have a pitch angle roughly 1.5 times larger than their midplane counterparts, however this exact value depends on the ratio of midplane to atmospheric temperature.

\section{Discussion}
\label{sec:discussion}

There is a considerable amount of azimuthal structure observed in the residual plots that extends over the bulk of the gas disk, far beyond the edge of the mm~continuum at $\sim 60$~au. In the following section, we characterise the observed spiral structures and discuss potential launching mechanisms.

\subsection{An Embedded Planet}
\label{sec:discussion:planet}

It known that embedded protoplanets will cause significant deviations in the 3D gas dynamics locally \citep{Kanagawa_ea_2015, Perez_ea_2015, Perez_ea_2018, Teague_ea_2018a}, with the observable signatures either a `kink' in the iso-velocity contours \citep{Pinte_ea_2018b, Pinte_ea_2019}, or a `Doppler-flip' around the planet \citep{Casassus_Perez_2019, Perez_ea_2019}. In this context, it is attractive to explain the perturbations in the velocity as due to an embedded proto-planet, a scenario explored in this section.

\begin{figure}
    \includegraphics[width=\columnwidth]{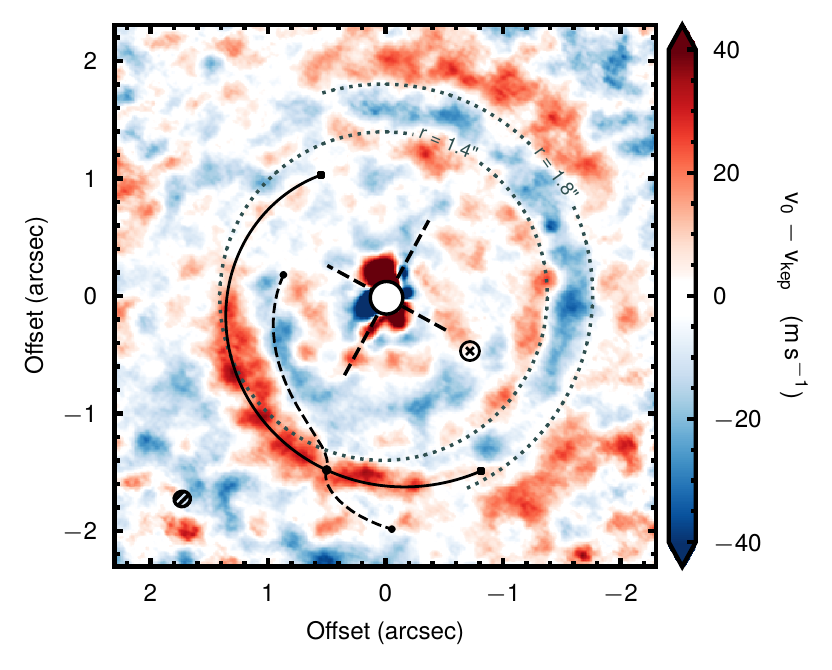}
    \caption{Annotated residuals from the $v_0$ map. A linear fit to the spiral is shown in the solid black line, while the dashed line shows a linear wake described by Eqn.~\ref{eqn:lindblad_phase} at the location of the planet proposed by \citet{Mentiplay_ea_2019}, with the orbital radius of 1.56\arcsec{} (94~au) The crossed region shows the location of the continuum excess reported by \citet{Tsukagoshi_ea_2019}. The dotted arcs show radii of 1.4\arcsec{} and 1.8\arcsec{} to highlight the radial structure. The major and minor axes are shown by the two perpendicular lines crossing the image center.}
    \label{fig:v0_residuals}
\end{figure}

Figure~\ref{fig:v0_residuals} shows a map of the residuals from the model $v_0$ maps with several features annotated. The two dotted lines show radii of 1.4\arcsec{} and 1.8\arcsec{}, showing that features in the NW are relative concentric, in contrast to the slight spiral of the feature in the SE. As the features we observe cross the major and minor axes of the disk, we are able to partially disentangle their direction.

As demonstrated in Appendix~\ref{sec:app:residuals_v0}, we expect deviations in $v_{\phi}$ to result in residuals that flip sign over the minor axis, deviations in $v_{\rm r}$ to result in residuals changing sign over the major axis, and perturbation in $v_{\rm z}$ to be constant as a function of azimuth. As the spiral extends over the disk major axis (PA = 151\degr{}), we can rule out significant $v_{\rm r}$ motions. Taking into account the projection of the velocities, a deviation of \ms{30} in $v_0$ relative to some background rotation corresponds to $\Delta v_{\rm z} \, / \, v_{\rm kep} \sim 1\%$ or $\Delta v_{\phi} \, / \, v_{\rm kep} \sim 10\%$. 

\citet{Pinte_ea_2019} show in their Supplementary Figures~5 -- 7 that an embedded planet will drive perturbations in all three directions, with the strength of the perturbations decreasing with height above the midplane for radial and rotational motions, and increasing with height for vertical. Given the face-on nature of TW~Hya, vertical motions, despite their intrinsically lower velocities relative to in-plane velocities, will be preferentially detected at high altitudes. The inferred velocities are broadly consistent with those predicted for embedded planets \citep{Szulagyi_ea_2014, Morbidelli_ea_2014, Fung_Chiang_2016, Dong_ea_2019}.

The arc in the SW of the disk crosses the minor axis of the disk so must be a significant radial or vertical velocity, again with similar magnitudes to the SE spiral. If this were vertical motions, we could be observing gas flowing towards the midplane, potentially due to gas falling into the recently opened gaps, again with a velocity of $\Delta v_{\rm z} \, / \, v_{\rm kep} \sim 1\%$.

Note that we will never be able to fully break the degeneracy between which velocity components are dominating the observed perturbations, however qualitative arguments such as the above can at least provide some guidance for the interpretation of the observations.

Given the comparable velocities of the azimuthal structure to simulations of embedded planets, we also compare the morphology of features. In addition to the Archimedean spiral fit described in the previous section, we attempt to fit the structures using a parameterization of a spiral wake from an embedded companion \citep{Bae_Zhu_2018a, Bae_Zhu_2018b},

\begin{equation}
\begin{split}
\phi(r) = \phi_p &-{\rm sgn}(r - r_p) \, \frac{\pi}{4m} \\
                 &-\left.\int_{r_m^\pm}^{r} \frac{\Omega(r')}{c_s(r')} \left| \left(1- \frac{r'^{3/2}}{r_p^{3/2}} \right)^2 - \frac{1}{m^2} \right|^{1/2} {\rm d}r' \right.
\end{split}
\label{eqn:lindblad_phase}
\end{equation}

\noindent where the companion is located at $(r_p,\, \phi_p)$ in disk midplane coordinates, $\Omega$ is the angular velocity and $c_s$ is the sound speed of the gas, $m$ is the azimuthal wave number, where the dominant component is given by $m = (1/2)(h/r)_p^{-1}$ and the limits for the integral are given by $r_{m}^{\pm} = ({ 1\pm 1/m})^{2/3} r_p$ which are the locations of the Lindblad resonances. As $m \rightarrow \infty$, we recover the linear limit described by \citet{Rafikov_2002} resulting in the most tightly wound (smallest pitch angle) spirals. We note that in the limit $r \gg r_p$ and $m \gg 1$, Eqn.~\ref{eqn:lindblad_phase} reduces to $\tan \beta \simeq (h / r) \times (r_p / r)^{1.5}$ such that, unless $T_{\rm gas} \propto r^2$, the pitch angle should decrease with radius.

For the location of a potential perturber, we adopt the location used in \citet{Mentiplay_ea_2019} at an orbital radius of 94~au and a polar angle of $\phi \approx 10\degr$. We use $m = 10^3$ in order to recover spirals with the smallest pitch angles and adopt $M_{\rm star} = 0.81~M_{\rm sun}$ to calculate $\Omega$ and using the radial $T_B$ profile to calculate a $c_s$ radial profile. The resulting profile is shown as the black dashed profile in Fig.~\ref{fig:v0_residuals}, which fails to reproduce the observed morphology.

A more tightly wound spiral was able to be generated but this required a $(h / r)_p \approx 0.02$, considerably smaller than the value used in \citet{Kama_ea_2016} to model the disk, $(h / r)_p \approx 0.1$. This suggests that if indeed the spirals were launched by an embedded protoplanet, the spirals are not generated by Lindblad resonances but rather by another mechanism, perhaps buoyancy resonance that excites intrinsically more tightly wound spirals \citep[][Bae et al., in prep.]{Zhu_ea_2012}.

Interestingly, \citet{Pinte_ea_2019} show that the morphology of $v_z$ motions vary from that of $v_r$ and $v_{\phi}$ motions driven by an embedded planet. In particular, $v_z$ motions manifest as a tighter spiral, centered on top of the embedded planet, rather than the in-plane motions which are known to flip signs either side \citep[c.f. the `Doppler flip';][]{Casassus_Perez_2019}. This morphology better matches the observations in the context of an embedded planet at a radius of $\sim 90$~au \citep{Teague_ea_2017, Mentiplay_ea_2019}. We note, however, that the simulations of \citet{Dong_ea_2019} show that there can be significant $v_z$ components significantly offset in the azimuthal direction from the embedded planet.

As shown in Fig.~\ref{fig:Tb_residuals}b, the velocity and temperature spirals appear to align, however do not fully overlap. It is unclear whether these are tracing the same underlying structure or two separate features. The similarity in pitch angle, Fig.~\ref{fig:Tb_residuals}c, and their location would favour the former scenario. One potential explanation would be that we are probing layers with different thermal properties. When we probe close to the disk surface where cooling is efficient, we would preferentially see spirals in the perturbed velocity. On the other hand, when we probe closer to the disk midplane, the intrinsic velocity perturbations would be small while the heat produced by spirals would be more efficiently trapped. The CO abundance in the TW~Hya disk is shown to decrease as a function of radius quite significantly \citep{Zhang_ea_2019} as well as a change in the slope of the $^{12}$CO $T_{\rm B}$ profile, indicative of a change in the optical depth of the transition \citep{Huang_ea_2018}, both supportive of this idea.

In order to distinguish between spiral launching scenarios it is essential that molecular emission spanning the full vertical extent are used to trace out the perturbations in the gas velocities and temperature. In particular, a prediction of the buoyancy resonances is that they generate perturbations which are strongest where the vertical temperature gradient is steepest which can easily be tested with a suite of optically thick and thin molecular tracers. Furthermore, higher spectral resolution data would enable a search for any locally broadened lines suggestive of large turbulent velocities which may betray an embedded planet \citep{Dong_ea_2019}.  

\subsection{Continuum Excess}
\label{sec:discussion:continuum}

Recently \citet{Tsukagoshi_ea_2019} reported the detection of continuum emission at a radius of $\approx 52$~au and a ${\rm PA} \sim 242\degr{}$. This location is marked in both Fig.~\ref{fig:Tb_residuals} and Fig.~\ref{fig:v0_residuals} as a circle with a cross inside. In neither of these residuals is any structure associated with this location seen. However, with a source size of $\approx 3$~au, it is likely that any structure would not be able to be resolved with our observations, which have a resolution of $\approx 8$~au, in addition to the $^{12}$CO emission tracing a vertical layer considerably higher in the disk than the midplane continuum emission.

\section{Summary}
\label{sec:summary}

We have used high angular resolution data of $^{12}$CO $J = 3-2$ emission to explore the physical structure of TW~Hya. Both the gas velocities and temperature structures show spiral structure when an azimuthally symmetric model is subtracted from the observations. Three dominant spirals are found, one in velocity and two in gas temperature, which span between 70 and 210~au and extend around the full azimuth of the disk. Despite the low significance of the residuals in an individual beam, the coherence of the structures over extended regions suggests real features. 

The spirals of warm gas bound a shadow observed in scattered NIR light \citep{vanBoekel_ea_2017}, consistent with the hypothesis that this is a region that is shadowed and thus receives less incident photons to heat the gas. The spiral in velocity overlaps the significant gap in the gas surface density at 90~au \citep{vanBoekel_ea_2017, Teague_ea_2017}, with perturbations of 10\% of the local Keplerian rotation if they are believed to be changes in the rotation speed. Equivalently, vertical motions on the order of 1\% of the rotation speed are also consistent.

All three spirals are described well by a linear (Archimedean) spiral with radially decreasing pitch angles, ranging from 9\degr{} at 70~au to 3\degr{} at 210~au. Tracing these spirals in the disk atmosphere through the optically thick $^{12}$CO emission suggests that the spirals at the midplane would be considerably more tightly wound \citep{Juhasz_Rosotti_2018}, inconsistent with most models of Lindblad-resonance driven spiral wakes \citep{Bae_Zhu_2018b, Bae_Zhu_2018a}.

In sum, these observations demonstrate a level of sub-structure in the gas hitherto unseen in protoplanetary disks. Detecting features in the velocity rules out features driven through chemical or excitation effects while correspondence with features observed in the scattered light further strengthens the idea of a dynamically active disk. 

\acknowledgments
We thank the anonymous referee who's comments improved the quality of the manuscript.
This paper makes use of the following ALMA data: ADS/JAO.ALMA\#2015.1.00686.S and 2016.1.00629.S. ALMA is a partnership of ESO (representing its member states), NSF (USA) and NINS (Japan), together with NRC (Canada), MOST and ASIAA (Taiwan), and KASI (Republic of Korea), in cooperation with the Republic of Chile. The Joint ALMA Observatory is operated by ESO, AUI/NRAO and NAOJ. R.T. and E.A.B. acknowledge funding from NSF grant AST-1514670 and NASA grant NNX16AB48G. J.B. acknowledges support from NASA grant NNX17AE31G. J.H. acknowledges support from the National Science Foundation Graduate Research Fellowship under Grant No. DGE-1144152.

\bibliography{main}{}
\bibliographystyle{aasjournal}

\appendix
\section{Rotation Map Fitting}
\label{sec:app:rotationmapfitting}

In this Appendix, we describe the fitting used to generate $v_{\rm mod}$. We use the \texttt{eddy} package to fit the $v_0$ map which was calculated, along with associated uncertainties, using \texttt{bettermoments}. The projected line centers are given by Eqn.~\ref{eq:vphi} where $\{x_0,\, y_0,\, i,\, {\rm PA} \}$ are used to deproject sky-plane coordinates $(x,\, y)$ into midplane cylindrical coordinates $(r,\, \phi)$ and $\{M_{\rm star},\, v_{\rm LSR}\}$ used to calculate the line center at a given location in the disk. This approach makes the implicit assumption that the rotation profile of Keplerian, and deviations due to radial pressure gradients or the self-gravity of the disk are negligible \citep{Rosenfeld_ea_2013, Pinte_ea_2018a}.

The fitting starts with an optimisation of the free parameters of using \texttt{scipy.optimize} which is then used as the starting positions for an MCMC exploration of the posterior distributions. The MCMC is performed using \texttt{emcee}, and utilises 256 walkers which take 500 steps to burn in, then an additional 10,\,000 steps to sample the posterior distribution function (PDF). The likelihood is calculated based on pixels between $r_{\rm min} = 0.26\arcsec{}$ (twice beam major axis) and $r_{\rm max} = 3.75\arcsec{}$ to limit the impact of pressure support most prominent in the outer disk. We then average 250 random samples to form the model rotation map, $v_{\rm mod}$. Percentiles of the posterior distributions are given in Table~\ref{tab:posteriors} where the uncertainties represent the (asymmetric) 16th to 84th percentile range about the median value.

\paragraph{Run 1}
For an initial attempt we allow both $i$ and $M_{\rm star}$ to vary, both with wide, flat priors spanning between $2\degr$ and $15\degr$ and $0.1~M_{\rm sun}$ to $2~M_{\rm sun}$, respectively. All other parameters, the source center, $(x_0,\, y_0)$, position angle, PA, stellar mass, $M_{\rm star}$, and systemic velocity, $v_{\rm LSR}$, were allowed to vary and assumed flat priors that extended far beyond any realistic values. All parameters other than $M_{\rm star}$ and $i$ rapidly converged with Gaussian PDF. $M_{\rm star}$ and $i$ were highly correlated as expected, but resulting in an $M_{\rm star} \sin i$ consistent with previous constraints \citep{Teague_ea_2016}.

\paragraph{Run 2}
We take the median inclination found in Run 1, $i = 5.8\degr$. The other values and priors were left as the same as before. This resulted in comparable PDFs, however with a much narrower distribution for $M_{\rm star}$ and $v_{\rm LSR}$.

\paragraph{Run 3}
In this run, we check to make sure that an elevated emission surface does not bias the result. We parameterize the surface as $z(r) = z_0 \cdot (r \, / 1\arcsec)^{\psi}$ and include a correction to $v_{\rm kep}$ to account for this. Flat priors were assumed for both $z_0 \in \{0, \, 0.5\}$ and $\psi \in \{0, \, 5\}$. Following \citet{vanBoekel_ea_2017} we assume the SW of the disk is closest to the observer. Neither $z_0$ nor $\psi$ converged with walkers spanning the whole range of priors. Due to the height correction to $v_{\rm kep}$, $M_{\rm star}$ converged to a slightly higher value then for the razor-thin disk.

\paragraph{Run 4}
As the tilt of the system (i.e. which side is of the disk is closer to the observer) is poorly constrained, we additionally tested the opposite scenario to Run 3, using a negative inclination of the same magnitude. This yield properties very similar to Run 3, however with a thinner, but more flared emission surface.\\

For all four runs, geometrical properties were found that were consistent (aside from $M_{\rm star}$). Similarly, all four runs found PDFs for the free parameters which resulted in an average standard deviation in $v_{\rm mod}$ of $\sim 1\%$, among both different samples, and different runs. For Runs 3 and 4 which considered elevated emission surfaces, it is likely that the structure observed in the residuals dominates the fit, rather than the difference in models. In the main text, we take $v_{\rm mod}$ generated using samples from Run 2.

\begin{deluxetable}{lcccccccc}
\tabletypesize{\footnotesize}
\tablewidth{0pt}
\tablecaption{Posterior distributions of Keplerian model parameters.}
\tablehead{
\colhead{Run} & \colhead{$x_0$} & \colhead{$y_0$} & \colhead{$i$} & \colhead{PA} & \colhead{$v_{\rm LSR}$} & \colhead{$M_{\rm star}$} & \colhead{$z_0$} & \colhead{$\psi$} \\
\coldhead{} & \colhead{$(\arcsec)$} & \colhead{$(\arcsec)$} & \colhead{$(\degr)$} & \colhead{$(\degr)$} & \colhead{$({\rm m\,s^{-1}})$} & \colhead{$(M_{\rm sun})$} & \colhead{$(\arcsec)$} & \colhead{-} 
}
\startdata
1 & $0.01 \pm 0.08$ & $0.01 \pm 0.08$ & $5.8_{-1.7}^{+4.0}$ & $151.3 \pm 6.6$ & $2840 \pm 23$ & $0.79_{-0.51}^{+0.77}$ & [0.0] & [1.0] \\
2 & $0.022 \pm 0.001$ & $0.023 \pm 0.001$ & [+5.8] & $151.4 \pm 0.1$ & $2841 \pm 1$ & $0.78 \pm 0.01$ & [0.0] & [1.0] \\
3 & $0.011 \pm 0.001$ & $0.017 \pm 0.001$ & [+5.8] & $151.4 \pm 0.1$ & $2841 \pm 1$ & $0.85 \pm 0.01$ & $0.176 \pm 0.006$ & $1.57 \pm 0.02$ \\
4 & $0.027 \pm 0.001$ & $0.026 \pm 0.001$ & [-5.8] & $151.4 \pm 0.1$ & $2841 \pm 1$ & $0.81 \pm 0.01$ & $0.055 \pm 0.003$ & $2.41 \pm 0.05$ \\
\enddata
\tablecomments{Uncertainties represent the 16th to 84th percentiles about the median value. These are the statistical uncertainties which do not take into account any systematic uncertainties arising from the choice of model which would likely be considerably larger. Values in square brackets were fixed during the MCMC.}
\label{tab:posteriors}
\end{deluxetable}

\section{Residuals in Rotation Maps}
\label{sec:app:residuals_v0}

\begin{figure}
    \centering
    \includegraphics[width=\textwidth]{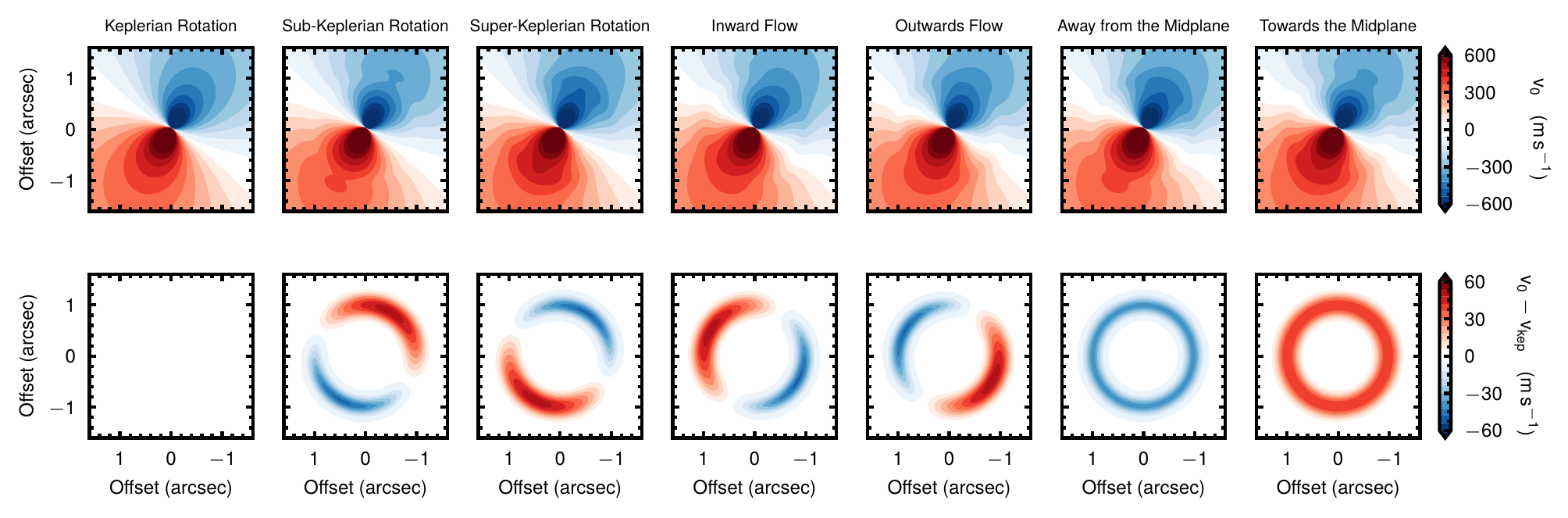}
    \caption{Analytical models of perturbations to the gas dynamics, top row, and their resulting residuals from a Keplerian model, bottom row.}
    \label{fig:analytic_models}
\end{figure}

Due to projection effects, residuals between maps of the line center, $v_0$, and some model rotation pattern, $v_{\rm mod}$, are non-trivial to interpret due to the projection terms going to zero along the axes. To demonstrate, in Fig.~\ref{fig:analytic_models} we show model rotation maps dominated by the background Keplerian rotation. To each model, we include a Gaussian shaped perturbation to the velocity field, centred at 1\arcsec{} with a standard deviation width of 0.2\arcsec{}. For the in-plane deviations ($v_{\phi}$ and $v_r$), we have a perturbation strength of 12\% of $v_{\rm kep}$, while for the out-of-plane deviations ($v_z$), these are an order of magnitude smaller at only 1.2\% of $v_{\rm kep}$. The bottom row are the residuals when subtracting the projected background $v_{\rm kep}$ map. Deviations in the plane, $v_{\phi}$ and $v_r$, result in residuals that flip sign across the minor and major axes, respectively. Conversely, the vertical motions, which have no projection term dependent on $\theta$, are constant as a function of azimuth. While it is possible to account for the projection with model data (dividing through by $\cos \theta$, for example), in practice this is not possible with real data as the noise will also be amplified significantly along the axes.

However, the difference in the spatial dependence of the projection terms for $v_{\phi}$, $v_{r}$ and $v_{z}$, allows us to disentangle these components. If residuals are seen to be constant across a given axis, then it is possible to rule out one potential component. For example, in Fig.~\ref{fig:v0_structure}b we see that the spiral is constant in sign across the minor axis. This suggests that this is not due to $v_r$ terms. Even if were a $v_r$ perturbation flipped sign close to the major axis, the lack of projection along the line of sight would mean that the residual would drop to zero.

While such an analysis will provide a guide, it is likely that in reality there are perturbations in all velocity components and it will be impossible to fully disentangle the three components. Forward modelling of hydrodynamical simulations promises a more accurate, however considerably slower, analysis.

\end{document}